

\input harvmac
\overfullrule=0pt
\def\Title#1#2{\rightline{#1}\ifx\answ\bigans\nopagenumbers\pageno0\vskip1in
\else\pageno1\vskip.8in\fi \centerline{\titlefont #2}\vskip .5in}

\baselineskip=18pt plus 2pt minus 2pt
%
%
\ifx\epsfbox\UnDeFiNeD\message{(NO epsf.tex, FIGURES WILL BE IGNORED)}
\def\figin#1{\vskip2in}
\else\message{(FIGURES WILL BE INCLUDED)}\def\figin#1{#1}\fi
\def\ifig#1#2#3{\xdef#1{fig.~\the\figno}
\goodbreak\midinsert\figin{\centerline{#3}}%
\smallskip\centerline{\vbox{\baselineskip12pt
\advance\hsize by -1truein\noindent\footnotefont{\bf Fig.~\the\figno:} #2}}
\bigskip\endinsert\global\advance\figno by1}
%
%
%
%

\def\ajou#1&#2(#3){\ \sl#1\bf#2\rm(19#3)}

\def\quart{{1 \over 4}}

\def\frac#1#2{{#1 \over #2}}

\def\hA{{\hat A}}
\def\hC{{\hat C}}

\def\eposq{{e^{2 \phi_0}}}
\def\epomsq{{e^{-2 \phi_0}}}
\def\dvdp{{\partial V \over \partial \phi}}
%
%
\lref\nappi{M. McGuigan, C. Nappi and S. Yost, \ajou Nucl. Phys. &B375
(92) 421.}
\lref\tbh{
E. Witten, \ajou Phys. Rev. &D44 (91) 314 \semi G. Mandal, A. Sengupta,
and S. Wadia, \ajou Mod. Phys. Lett. &A6 (91) 1685.}
\lref\CGHS{C.G. Callan, S.B. Giddings, J.A. Harvey, and A. Strominger,
``Evanescent black holes,"\ajou Phys. Rev. &D45 (92) R1005.}
\lref\BDDO{T. Banks, A. Dabholkar, M.R. Douglas, and M O'Loughlin, ``Are
horned particles the climax of Hawking evaporation?'' \ajou Phys. Rev.
&D45 (92) 3607.}
\lref\corn{T. Banks and M. O'Loughlin, ``Classical and quantum production
of cornucopians at energies below $10^{18}$ GeV'', Rutgers
preprint RU-92-14 (1992).}
\lref\BD{C. Brans and R. H. Dicke, \ajou Phys. Rev. &124 (61) 925.}
\lref\BR{B. Bertotti, \ajou Phys. Rev. &116 (59) 1331 \semi I. Robinson, \ajou
Bull. Acad. Polon. Sci. &7 (59) 351.}
\lref\GS{S. Giddings and A. Strominger, \ajou Phys. Rev. &D46 (92) 627.}
\lref\growhair{S. Coleman, J. Preskill, and F. Wilczek, ``Growing
Hair on Black Holes,'' \ajou Phys. Rev. Lett. &67 (91) 1975.}
\lref\qhair{S. Coleman, J. Preskill, and F. Wilczek, ``Quantum Hair
on
Black Holes'' IASSNS-HEP-91/64, CALT-68-1764, HUTP-92/A003, (1991)
and
references therein.}
\lref\dgt{F. Dowker, R. Gregory, and J. Traschen, ``Euclidean Black
Hole
Vortices,''
FERMILAB-Pub-91/142-A, EFI-91-70 (1991).}
\lref\bghhs{M. J. Bowick et. al.}
\lref\PSSTW{J. Preskill, ``Quantum hair'', CALT-68-1671 (1990);
J. Preskill, P. Schwarz, A. Shapere, S. Trivedi and
F. Wilczek, ``Limitations on the statistical description of black
holes,''
IAS preprint IASSSNS-HEP-91/34.}
\lref\wittenbh{E.Witten, \ajou Phys. Rev. D &44 (91) 314.}
\lref\bghhs{M.Bowick, S.Giddings, J.Harvey, G.Horowitz and
A.Strominger, \ajou Phys. Rev. Lett. &61  (88) 2823.}
\lref\kw{L.Krauss and F.Wilczek, \ajou Phys. Rev. Lett. &62 (89)
1221.}
\lref\hs{ For a review see J. Harvey and A. Strominger, ``Quantum Aspects of
Black Holes,''
preprint EFI-92-41, hep-th/99209055.}
\lref\GM{G. Gibbons and K. Maeda, \ajou Nucl. Phys. &B298 (88) 741.}
\lref\GHS{D.Garfinkle, G.Horowitz and A.Strominger, \ajou Phys. Rev. &D43
 (91) 3140.}

\Title{\vbox{\baselineskip12pt
\hbox{EFI-92-49}\hbox{hep-th/9209070} }}
{\bf BLACK HOLES WITH A MASSIVE DILATON}
\bigskip
\centerline{Ruth Gregory and Jeffrey A. Harvey}
\bigskip
\centerline{\it Enrico Fermi Institute, University of Chicago}
\centerline{\it 5640 Ellis Avenue, Chicago IL 60637}
\bigskip
\centerline{\bf Abstract}
The modifications of dilaton black holes which result when
the dilaton acquires a mass are investigated. We derive
some general constraints on the number of horizons of the
black hole and argue  that if the product of the charge
$Q$ of the black hole and the dilaton mass $m$ satisfies
$Q m < O(1)$ then the black hole has only one horizon. We also
argue that for $Q m > O(1)$ there may exist solutions with three
horizons and we discuss the causal structure of such solutions.
We also investigate the possible structures of extremal solutions
and the related problem of two-dimensional dilaton gravity
with a massive dilaton.
\bigskip
\medskip
\Date{9/92}
\eject
\newsec{Introduction}

The notion that Einstein's theory of gravity should be modified
by the addition of scalar fields has a long history dating
back to the pioneering work of Brans and Dicke \BD\ who were
motivated by the desire to more directly incorporate Mach's
principle into physical law. In recent times a particular
variant of this idea, dilaton gravity, has received attention
because of its close connection with low-energy string theory.
In this theory neutral black holes  are still described
by the Schwarzschild metric and the scalar dilaton plays no role.
For charged black holes however the dilaton plays a crucial role
in modifying the causal structure of the solution.

The causal structure of charged black holes described
by the Reissner-Nordstr{\o}m metric in Einstein
gravity has given rise to a number
of puzzles and speculations. The solution has an outer event
horizon at $r_+$, a Cauchy horizon at $r_-$, and a timelike
singularity in place of the spacelike singularity of the
Schwarzschild solution. This gives rise to many peculiar features.
For example, an observer crossing the Cauchy horizon at $r_-$ would
see the whole history of the asymptotically flat region she
originated in flash by in finite time at infinite blue shift and
then find that her future is no longer determined by her past.
These bizarre features suggest that the inner Cauchy horizon
is unstable against small perturbations. In contrast, the
charged dilaton black hole has a Schwarzschild type causal structure
with only one horizon, and a spacelike singularity, suggesting
stability of the solution to perturbations.

In a different direction, there have been recent attempts
to unravel the mysteries of Hawking radiation in a class of
two-dimensional theories of dilaton gravity \hs. These theories
can be viewed as low-energy effective theories of four-dimensional
extremal dilaton black holes. In contrast to the extremal
Reissner-Nordstr{\o}m black hole, the extremal dilaton black hole
has the singularity and horizon merging at ``$r=2M$'' and the
horizon actually becomes an internal scri of the
spacetime, thus giving the four-dimensional spacetime the
causal structure of the two-dimensional linear dilaton
vacuum reviewed in \hs. In some attempts to understand resolutions
to
the puzzles raised by Hawking radiation the infinite
``throat'' structure of extremal dilaton black holes has played a
significant
role \refs{\CGHS, \BDDO, \corn}.
The idea here is that the infinite volume of the
throat can store the arbitrarily large amount of information
which may be lost in the standard semi-classical picture
of Hawking evaporation of a black hole.

So far the structure of dilaton black holes
is understood only in the physically unrealistic limit of
vanishing dilaton mass. If string theory and its low-energy
limit are relevant to the real world
then the dilaton must eventually acquire a mass.
We would then like to know how the above features are modified
by the presence of a dilaton mass. In particular, we would
like to know what causal structures are allowed and whether
the feature of an infinite ``throat'' persists.
Unfortunately
our current understanding of how the dilaton acquires
a mass is rather
primitive and is tied to our lack of understanding of supersymmetry
breaking.
Since we do not have a good model of
how the dilaton mass is generated, we perform as much of the analysis
as possible for a general choice of dilaton potential and when we
need an explicit choice of potential we consider
two simple choices of mass term which we hope will reflect
the general structure of such solutions.

The outline of the paper is as follows. The second section contains
a review
of massless dilaton black holes, and serves to establish our notation
and conventions.
In section three we discuss adding a mass term, derive general
constraints
on the number of horizons, show that there is only one horizon when
$Q m < O(1)$ and derive expansions for the behavior of the solutions
in various asymptotic regions.
In section
four we discuss the structure of the possible extremal limits
of massive dilaton black holes.
Section five contains a brief discussion
of two-dimensional massive dilaton gravity.
We end with some brief final
comments in section six.

\newsec{Massless dilatonic black holes}

Black holes in  dilaton gravity were first
analyzed in some generality by Gibbons and Maeda \GM. An
elegant form of the solution was given in later work by
Garfinkle Horowitz and Strominger (GHS) \GHS\ and we will for
the most part follow their approach.
GHS considered a massless dilaton
field coupled to electromagnetism and gravity. Taking the signature
of the metric to be ($+,-,-,-$) the appropriate action is
\eqn\action{
S=\int d^4x \sqrt{-g} [ -R + 2(\nabla\phi)^2 - e^{-2\phi} F_{ab}^2 ],
}
and one wants to find static, spherically symmetric solutions
with non-trivial dilaton field.
The metric may be written in the general form
\eqn\metric{
ds^2 = A^2(r) dt^2 - A^{-2}(r) dr^2 - C^2(r) (d\theta^2 +
\sin^2 \theta d\phi^2)
}
where $A(r_+)=0$ marks the outermost event horizon,
and $C(r_+)^2 = {\cal A}/4\pi$ is given in terms
of the area of the event horizon. The Hawking temperature,
$\beta^{-1}$, of the black hole is given by
$(A^2)'|_{r_+} = 4\pi \beta^{-1} $.

Varying the action \action\ gives the equations of motion
\eqn\max{
\nabla_a [ e^{-2\phi} F^{ab} ] =0
}
\eqn\dilly{
\nabla_a \nabla^a \phi = \half e^{-2\phi} F_{ab}^2
}
and the ``Einstein'' equations
\eqn\einst{
\eqalign{
G^0_0 &= {1\over C^2} (1-A^2 C'^2) - {2AA'C' \over C}
- {2A^2 C'' \over C} = {\cal T}^0_0 \cr
G^r_r &= {1\over C^2} (1-A^2 C'^2) - {2AA'C' \over C}
= {\cal T}^r_r \cr
G^\theta_\theta &= - \half (A^2)'' - {2AA'C' \over C}
- {A^2 C'' \over C} = {\cal T}^\theta_\theta \cr
}
}
with
\eqn\em{
{\cal T}_{ab} = 2\nabla_a \phi \nabla_b \phi - 2 e^{-2\phi}
F_{ac} F_b^{~c} - \half g_{ab} [ 2 (\nabla \phi)^2 -
e^{-2\phi} F_{cd}^2 ].
}
Notice that the field strength of a magnetically charged black hole
in Einstein theory, $F = Q\sin \theta d\theta \wedge d\phi$, also
satisfies the equation of motion \max\ for dilaton gravity. However,
if there is a non-zero electromagnetic field strength then \dilly\
demands a non-trivial dilaton field. This is to be contrasted with
zero electromagnetic field where a no-hair theorem demands that the
dilaton vanish.

Looking for
a static monopole solution, and re-arranging the equations of motion
yields:
\eqn\zerom{
\eqalign{
C'' &= - \half {C\over A^2} ({\cal T}^0_0 - {\cal T}^r_r) = -C\phi'^2
\cr
((A^2)'C^2)' &= - C^2( 2{\cal T}^\theta_\theta + {\cal T}^r_r
- {\cal T}^0_0) = {2Q^2 e^{-2\phi} \over C^2} = -2 (A^2C^2\phi')' \cr
(A^2C^2)'' &= 2 - 2C^2  ({\cal T}^\theta_\theta + {\cal T}^r_r) = 2
\cr
}}
The last two of these equations are readily integrated to yield
\eqn\fat{
\eqalign{
(A^2)'C^2 + 2A^2C^2\phi' &= {\cal A}/\beta \cr
A^2C^2 &= (r-r_+)^2 + {{\cal A}\over \beta} (r-r_+) \cr
}}
{}From this, it is straightforward to show that choosing
$\beta = 4\pi r_+ =8\pi M$
\eqn\ghsoln{
\eqalign{
A^2&= 1 - {r_+ \over r} \cr
C^2 &= r(r-{2Q^2e^{-2\phi_0} \over r_+} ) \cr
e^{-2 \phi} &= e^{-2 \phi_0} C^2/r^2  \cr
}}
where $\phi_0$ is the value of the dilaton field at infinity. One
obtains an electrically charged solution from the magnetic solution
above by performing a  modified duality transformation on the
electromagnetic
field and changing the sign of the dilaton. Starting from the field
strength, dilaton, and metric for a magnetic solution, $F^M$, $\phi^M$,
and $g^M$ we obtain the electric solution as
\eqn\dual{\eqalign{F_{\mu \nu}^E &  = \half e^{-2 \phi^M}
          {\epsilon_{\mu \nu}}^{\lambda \rho} F_{\lambda \rho}^M \cr
          \phi^E  & = -\phi^M \cr
	  g_{\mu \nu}^E & = g_{\mu \nu}^M . \cr }}

The important differences to stress between the dilaton and Einstein
magnetic black holes are the horizon/singularity structure and the
nature of the extremal limit. The dilaton black hole has only one
horizon and a spacelike singularity, giving rise to a Schwarzschild
type
Penrose-Carter diagram; on the other hand the typical Einstein magnetic
black
hole has two horizons and a timelike singularity, giving rise to the
familiar `vertically-infinite' Penrose-Carter diagram for the
Reissner-Nordstr{\o}m solution shown in fig. 1.
\ifig\fone{Penrose-Carter diagram  for a Reissner-Nordstr{\o}m black
hole.}{\epsfysize=4.8in \epsfbox{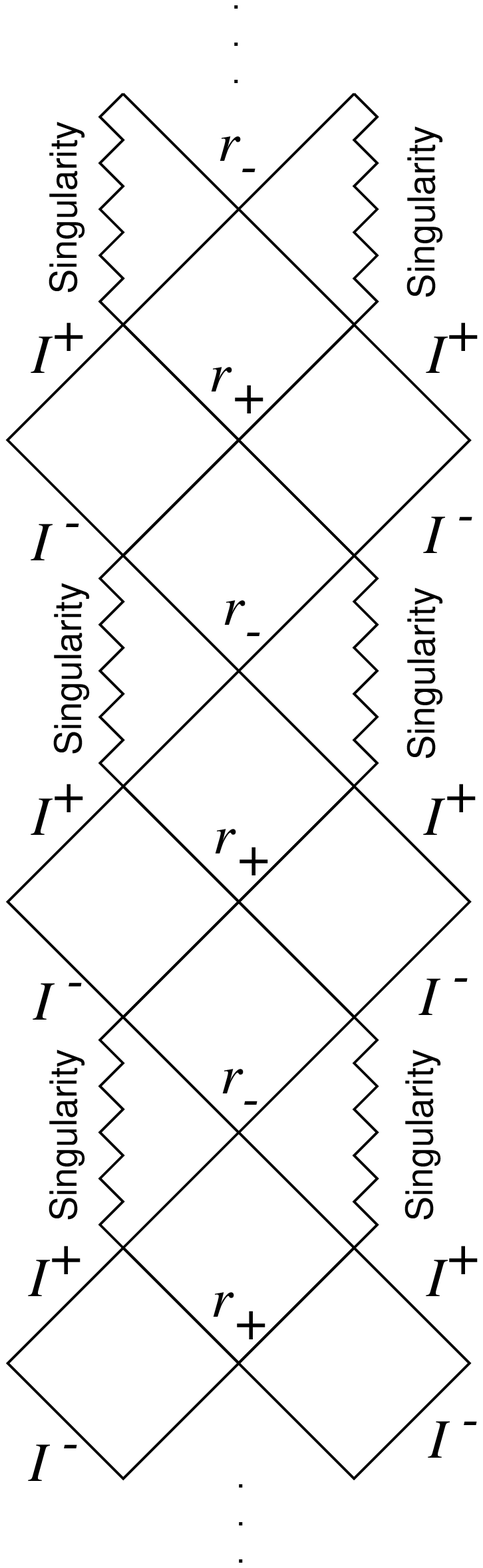}}
The extremal limits of these two types of black hole differ both in
their charge-mass ratios and in their structure. For the dilaton
hole the extremal limit is
$Q^2=2M^2 e^{2 \phi_0}$, as opposed to $Q^2=M^2$ in the
Reissner-Nordstr{\o}m case.
Further, the event
horizon is actually singular in this limit and has zero area, as
opposed
to the Reissner-Nordstr{\o}m case which has finite area and is
non-singular.

It is also interesting to analyze the solution in terms of the
string metric ${\hat g}_{ab}$ defined by
\eqn\stringg{
{\hat g}_{ab} = e^{2\phi} g_{ab}.
}
The introduction of ${\hat g}$ is motivated in part by the fact that
in fundamental string theory the string world-sheet has minimal
surface area with respect to the metric ${\hat g}_{ab}$. The
line element is then
\eqn\linel{d {\hat s}^2 = {(1-2M e^{\phi_0}/r) \over
                           (1-Q^2 e^{-\phi_0}/Mr)} dt^2 -
                          { dr^2 \over
                          (1-2M e^{\phi_0}/r)(1-Q^2 e^{-\phi_0}/Mr)}  -
                          r^2  d \Omega_{I\negthinspace I}^2 }
where we have absorbed the factor of $e^{2 \phi_0}$ in order that the
metric have the canonical asymptotic form.
Note that in this metric the singularity at $r_s=Q^2 e^{- \phi_0}/M$
corresponds to a two-sphere of area $4 \pi Q^4 e^{-2 \phi_0}/M^2$
rather
than to a point as in the Einstein metric.

In the extremal limit $Q^2 \rightarrow 2M^2 e^{2 \phi_0}$ the
line element \linel\ becomes
\eqn\linelim{d {\hat s}^2 =  dt^2 - {dr^2 \over (1-2M/r)^2}
                           -r^2 d \Omega_{I\negthinspace I}^2 .}
In this limit the previous singularity at $r_s$ coincides with
the horizon at $r_+=2M e^{\phi_0}$ and both have been pushed off to
infinite
proper distance. In terms of a new coordinate, $\sigma$, with
\eqn\sigcoor{d \sigma = {dr \over (1-2M/r)} ,}
we have as $r \rightarrow 2M$
\eqn\tube{d {\hat s}^2 \rightarrow  (-dt^2 + d \sigma^2
            +(2M)^2 d \Omega_{I\negthinspace I}^2 )}
so that the geometry approaches that of an infinite ``tube''
of radius $2M e^{\phi_0}$.

However, note that this infinite tube is quite distinct from
the infinite tube of the spatial sections of extremal
Reissner-Nordstr{\o}m black holes. In this extremal metric,
$r=2M=\sqrt{2} Q$  is
not
only at an infinite spatial distance, but also at an infinite proper
distance
to any causal observer so that in effect the event horizon provides
another internal asymptotic null infinity as can be seen from the
Penrose-Carter
diagram of fig. 2.
\ifig\ftwo{Penrose-Carter diagram of an extremal dilaton black hole. The
subscripts
$AF$ and $Th$ refer to the asymptotically flat  and  ``throat''
regions of the black hole respectively.}
{\epsfysize=4.8in \epsfbox{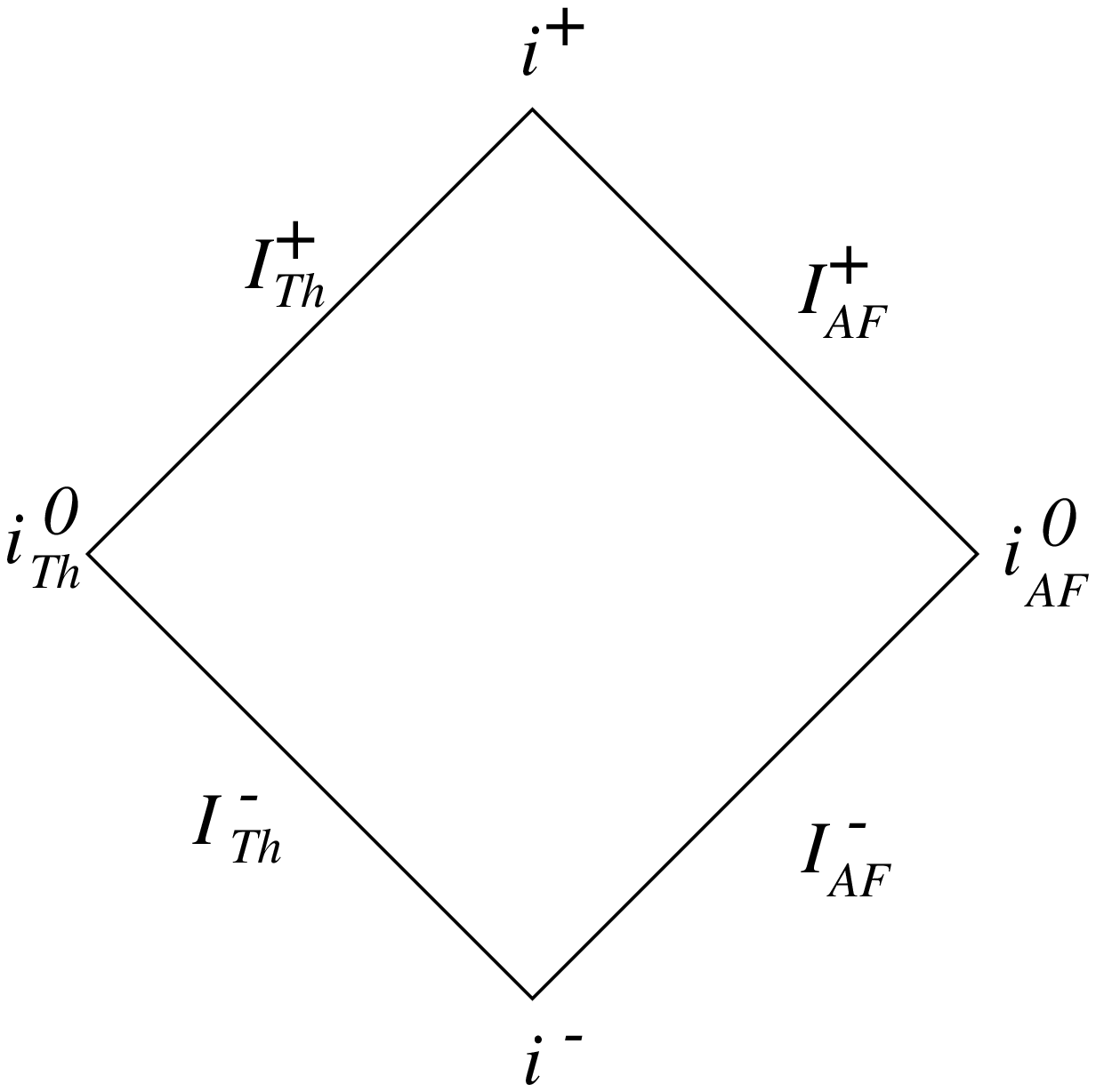}}
For extremal Reissner-Nordstr{\o}m however, while
$r=M=Q$ is located at the end of an infinite tube
in a spatial section of the metric, it is at a finite proper
distance for any infalling observer--thus there is no internal
asymptotic region, only an event horizon as shown in fig. 4.
It is also worth
emphasizing that the infinite throat of the extremal solution in
the string metric occurs only for the
the magnetically charged solution and not for the electrically
charged solution.

In the following section we will also find useful an
alternative parametrization
of the string metric in the form \metric. We can write
\eqn\sthat{d {\hat s}^2 = \hA^2 (\rho) dt^2 - \hA^{-2} (\rho)
d \rho^2 - \hC^2 (\rho) d \Omega_{I\negthinspace I}^2}
with
\eqn\hatdefs{\eqalign{ \hC^2(\rho) &= r^2(\rho) \cr
             \hA^2(\rho) &= {(1-2M/r(\rho))\eposq \over
             1-Q^2 \epomsq /Mr(\rho) } \cr }}
with $r(\rho)$ defined implicitly by
\eqn\rhodef{ \rho = \eposq(r-r_0) + {Q^2 \over M}
             \log \left( {r-Q^2 \epomsq /M \over r_0 -Q^2 \epomsq /M}
                  \right)}
and $r_0$ the (arbitrary) point at which $\rho=0$. Note that
the singularity $(r=r_s)$ occurs at $\rho=-\infty$ and that
$\rho \rightarrow \eposq r$ as $r \rightarrow \infty$.

In this metric the equations of motion can be seen to admit solutions
which are products of two two-dimensional solutions. In particular,
they admit a solution of the form $S^2 \times {\cal M}_{BH}^2$ where
$S^2$ is a two-sphere of constant radius and ${\cal M}_{BH}^2$ is
a two-dimensional black hole solution. Explicitly this solution
is given by
\eqn\decomp{\eqalign{ {\hat C}^2 (\rho) &= 2 Q^2 \cr
                      {\hat A}^2 (\rho) &= 1 - 2M e^{-2 \lambda \rho} \cr
		      \phi &= -\lambda \rho \cr}}
where $\lambda^2 = 1/(8 Q^2)$ and $M$ is the arbitrary mass of the
two-dimensional black hole. For a more detailed explanation of the relation
between four- and two-dimensional dilaton black holes
see \GS. We will see later that the addition of a dilaton potential no longer
allows solutions of the above form.

To summarize: in either metric the important features to note are
that the equations
of motion are readily solved, the horizon structure of a dilaton
black hole is different than in Einstein gravity, and the
thermodynamical
relationships are also different. These latter features will persist
when we
add a mass term for the dilaton, although the equations will not be
so
easy to solve!

\newsec{Massive dilatonic black holes}

In this section we consider adding in a potential term for the
dilaton field. Instead of the value of the dilaton at infinity and
hence the string coupling constant being arbitrary, it will
now be determined by the minimum of the potential.
Since we want to generate a mass term for
the dilaton, the leading term in the potential should
be $V_1(\phi)=2 m^2 ( \phi - \phi_0)^2$ (the factor of two is due to
the
unconventional normalization of the kinetic term in \action).
In string theory the natural
variable is $e^{\phi}$ which plays the role of the
dimensionless coupling constant. We thus expect that the
true potential for the dilaton will also have higher
order corrections in an expansion in $\phi$. Where possible,
we will try to make general statements without making
detailed assumptions about the form of the potential. When
we need to consider specific possibilities we will consider
either the potential $V_1$ or $V_2 =2 m^2 \sinh^2 (\phi - \phi_0)$.
This latter choice
of potential is a simple function of $e^{\phi}$
which agrees with $V_1$ to lowest order in $\phi^2$ but
which is more divergent at the singularity of the black
hole than $V_1$.

Intuitively, we expect that if a field has mass $m$, then at length
scales large compared to $m^{-1}$
the potential will suppress fluctuations in the
field while at
lengths small compared to $m^{-1}$ it will behave rather like a
massless field. Therefore we would expect that at large distances our
black hole would look like a Reissner-Nordstr{\o}m black hole, and at
small distances like a massless dilaton black hole.
We might also expect that the classical treatment here
will break down unless
the Compton wavelength of the dilaton $1/m$ is small compared
to the gravitational radius of the black hole $\sim M/M_P^2$ with
$M_P$ the Planck mass. In geometrized units this means
$m M \gg 1$.

For large black holes (i.e. those satisfying this
criterion) we thus expect the
structure to be asymptotically Reissner-Nordstr{\o}m.
If the black hole has large charge $Q \sim M$ then both
the outer and inner horizons of the Reissner-Nordstr{\o}m solution
occur in a region where we expect the dilaton to play
a negligible role. However, as we
approach the singularity a new
possibility arises. Namely that the dilaton ``switches on'' and
the solution becomes like the massless dilaton black hole
which would cause a third
horizon and the causal lattice shown in fig 3.
\ifig\fthree{Penrose-Carter diagram for a massive dilaton black hole with
outer horizon, $r_+$, middle horizon $r_0$, and inner horizon $r_-$.
${\it I}^\pm$ indicate future and past null infinity respectively,
while the shading indicates regions not included in the spacetime.
The figure repeats periodically to tile the plane.}
{\epsfysize=4.8in \epsfbox{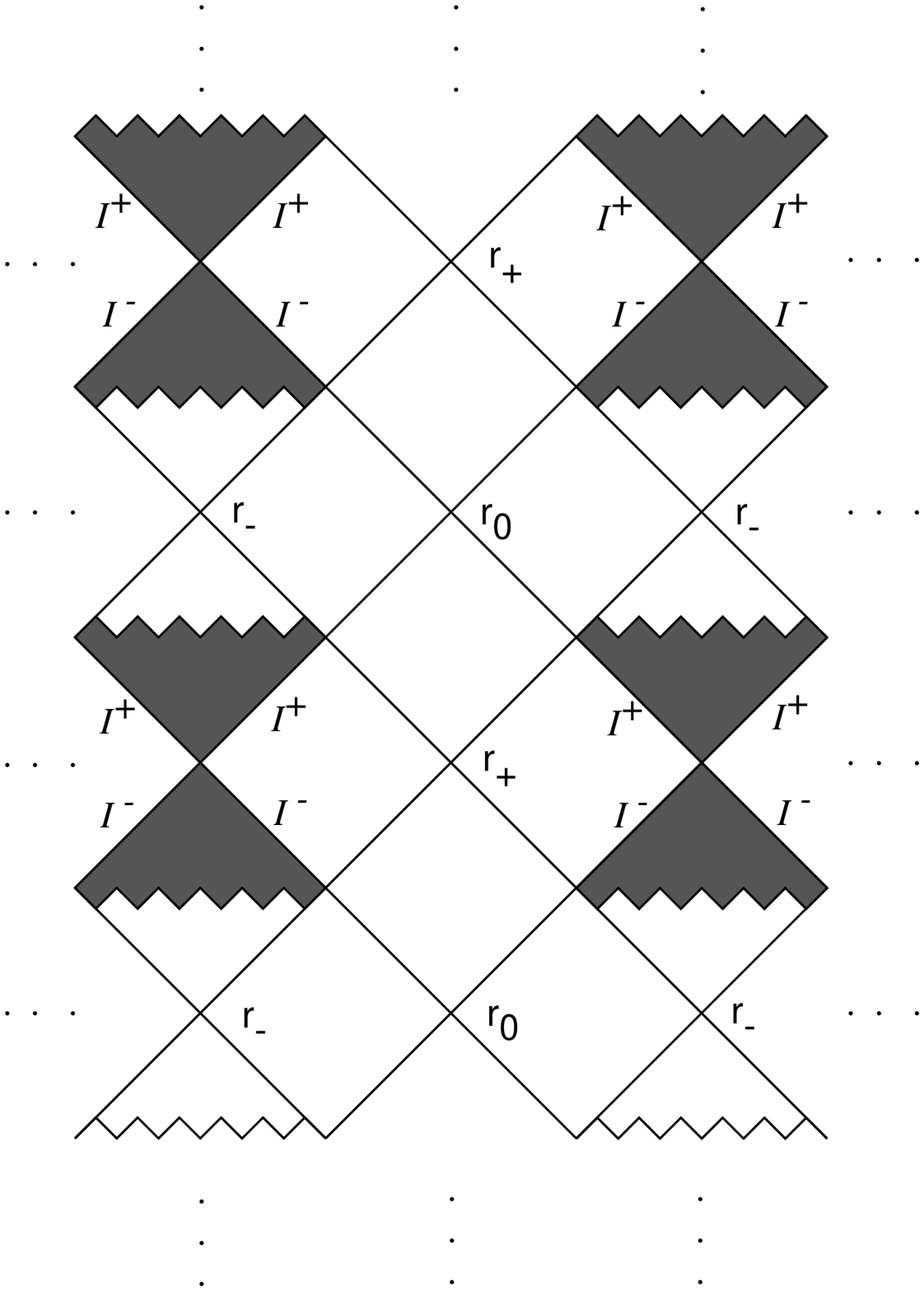}}
As we will see,
this final possibility requires a violation of the strong energy
condition, which in turn requires the dilaton to become dominant at
some length scale - thus validating our intuition. On the
other hand, for real physical black holes we should replace the
magnetic charge $Q$ by the electric charge,  $Q_e$,
and we would expect $Q_e \ll 10^{18} M$ since otherwise
the black hole would neutralize itself by attracting charged
particles from the surrounding medium. In this case  we expect the
dilaton to become important long before we reach the inner horizon
of the Reissner-Nordstr{\o}m solution and the exact solution may have
one, two, or three horizons.

The organization of this section will be as follows: We first set up
the equations of motion in the Einstein metric and discuss some
general properties of solutions, examining under what circumstances
three horizons {\it may} occur. We then obtain a specific restriction
on the number of horizons, given a certain constraint which is
satisfied by our test potential $V_2$, but not $V_1$. We then examine
the equations of motion in the string metric, giving a plausibility
argument that for $Qm<O(1)$ the solution will have only one horizon.
We conclude by commenting on the dual electric solutions. Note that we
will be concerned only with the non-extremal black holes in this section.

We begin by stating the general equations of motion.
Remaining with the general expression $V(\phi)$, we
note that the action becomes
\eqn\maction{
S=\int d^4x \sqrt{-g} [ -R + 2(\nabla\phi)^2 - V(\phi) - e^{-2\phi}
F_{ab}^2 ]
}
Thus the Maxwell equations are unchanged, and the
equations of motion and energy momentum are modified to
the following:
\eqn\mdilly{
\nabla_a \nabla^a \phi + \quart {\partial V \over \partial \phi}
= {Q^2 e^{-2\phi} \over C^4}
}
\eqn\mem{
{\cal T}_{ab} = 2\nabla_a \phi \nabla_b \phi - 2 e^{-2\phi}
F_{ac} F_b^{~c} - \half g_{ab} [ 2 (\nabla \phi)^2 -
e^{-2\phi} F_{cd}^2 - V(\phi) ]
}
which implies:
\eqn\massv{
\eqalign{
C'' &= -C\phi'^2 \cr
((A^2)'C^2)' &= C^2(\half{\partial V\over \partial\phi} - V) -2
(A^2C^2\phi')' \cr
(A^2C^2)'' &= 2 - 2 C^2 V(\phi). \cr
}}
Note that the presence of the potential means that we no longer have
a
first integral to simplify the process of solving these equations. A
few general remarks however can be made.

Note that the first of the equations \massv\ implies that $C'$ is
always decreasing, therefore if we wish
to have $C'(r) \to 1$ at infinity, then, as with the massless case,
the singularity will occur at positive $r$, $r_{\rm sg}>0$. (In fact
the dominant energy condition guarantees $r_{\rm sg} \geq 0$.)
The final equation of \massv\ shows that with a potential, $(A^2C^2)''$
is no longer necessarily positive. Since, roughly speaking, it is the
positivity of $(A^2C^2)''$ that guarantees a single horizon, we see
that if $C^2V>1$, then multiple horizons are possible.
The middle equation of \massv\ actually gives us some more concrete
requirements on the stress-energy-momentum tensor, and hence
$V(\phi)$, if there are to be three horizons.
Note that the existence of three horizons requires
(at least) two zeros of $(A^2)'$. Since $((A^2)'C^2)' =
-C^2 (2 {\cal T}^\theta_\theta + {\cal T}^r_r -{\cal T}^0_0)$
this in turn implies a violation of the strong energy condition (SEC).
A violation of the SEC is by no means impossible,
for although electromagnetism satisfies the strong energy condition,
a massive scalar field does not necessarily, nonetheless, it again
emphasizes the point that if the scalar potential is weak, the horizon
structure of the massless dilaton black hole will not be altered.

The dilaton equation, \mdilly, can be used to prove that if the
potential $V(\phi)$ is convex, then $\phi' <0$ when $A^2 >0$.
To show this we first argue that $\phi'<0$ for some range outside
an event horizon, and then for the whole range.

Multiplying \mdilly\ by $\phi'$ and integrating yields
\eqn\intdil{\eqalign{ \int {Q^2 e^{-2 \phi} \over C^4 } \phi' &=
\int \half \phi' \dvdp - {\phi' \over C^2} (A^2 C^2 \phi')' \cr
{}&= \half [ V-A^2 {\phi'}^2 ] -
\int ({2 C' A^2 \over C} +\half (A^2)' ){\phi'}^2 . \cr }}
Evaluating this integral between the horizon and infinity shows
\eqn\showthat{ \int_{r_+}^\infty {Q^2 e^{-2 \phi} \over C^4} \phi' <
0}
{\it i.e.} that $\phi' <0$ at some point.

Now that we have shown that $\phi'<0$ at some point outside
the horizon we will show that this holds at all points outside
the horizon. The proof is by contradiction. If $\phi$ is not
monotonically decreasing outside the horizon then
it must have a maximum outside the horizon (possibly at infinity).
At this maximum \mdilly\ then implies
\eqn\mdilimp{ \half \dvdp - {Q^2 e^{-2 \phi} \over C^4} = A^2 \phi''
\le 0,}
(with equality for the case where the maximum is at infinity).
As we move from this maximum in towards the horizon $\phi$
can  either have a minimum or decrease to the horizon.
Equation \mdilly\ then implies that
\eqn\prochoice{ \half \dvdp - {Q^2 e^{-2 \phi} \over C^4} =
                \phi'' A^2 + {(A^2)}' \phi' + {2 C' A^2 \phi' \over
C} >0.}
The r.h.s.\ of \prochoice\ is positive for either case
since at the supposed minimum
$A^2$ and $\phi''$ are positive
and $C$ is monotonically increasing while in the latter
case the r.h.s.\ of \prochoice\ is positive at the horizon
since $A^2=0$ and $(A^2)' \phi' >0$. So we have established
that the l.h.s.\ of \prochoice\ must be positive at or outside
the horizon if $\phi'$ is positive at or outside the horizon.
However, for $\phi' >0$ and a convex potential, the l.h.s.\
of \prochoice\ is an increasing function of $r$. Hence
it cannot become non-negative given \mdilimp. We have
thus derived a contradiction and shown that $\phi'$
remains negative everywhere outside the event horizon.
A similar argument shows that we expect $\phi'<0$ in any inner
regions of the black hole where $A^2 >0$.

Having used the equations to extract some general information,
let us now be specific about the form of the solution.
We can confirm some of the previous reasoning by solving the
equations  \massv\ in a power series in $1/r$, that is at large
distances from the singularity. To lowest non-trivial order the
asymptotically flat solution depends only on the quadratic part
of the potential and therefore will essentially be
independent of the potential, being given by
\eqn\rnexp{\eqalign{ \phi(r) &= \phi_0 + {Q^2 e^{-2 \phi_0}
                      \over m^2 r^4} +
                     \cdots +  F e^{-mr} +  \cr
                     A^2(r) &= 1-{2M \over r} + {Q^2 e^{-2 \phi_0}
		                \over r^2} -
                                {Q^4 e^{-4 \phi_0} \over 5 m^2 r^6}
				+ \cdots \cr
                     C(r) &= r - {2 Q^4 e^{-4 \phi_0}
		           \over 7 m^4 r^7} + \cdots \cr
}}
where $F$ is an arbitrary constant.
This asymptotic expansion agrees with the Reissner-Nordstr{\o}m
solution
at large distances, remembering that $\phi_0$ shifts
the value of the gravitational coupling at infinity,
and indeed up to $r^2 \sim e^{- \phi_0}Q/m$, which will be
past at least the event horizon for large mass black holes.
In addition, note that $\phi$ is monotonically decreasing as claimed.

Examining the equations of motion near the
central singularity ($C^2\to0$), the form of the potential becomes
important. For our two test potentials, $V_1$ and $V_2$,
we find the following behavior for the metric and the dilaton
\eqn\origin{
\eqalign{
C &\sim c_0  (r-r_0)^{1/2} + c_1 (r-r_0)^{3/2} \cr
e^{-2\phi} &\sim f_0 (r-r_0) - {2c_1 \over c_0} f_0 (r-r_0)^2 \cr
A_1^2 &\sim {1-2f_0Q^2/c_0^2 \over c_0 c_1} + {2f_0 Q^2 \over c_0^4}
(r-r_0) \cr
A_2^2 & \sim A_1^2 - {m^2 (r-r_0) \over 2 f_0} \cr}}
where $A_i^2$ is the $g_{00}$ appropriate for $V_i$. For $V_1$,
$(A^2)'$ is always positive at the singularity, whereas for $V_2$ the
sign of $(A^2)'$ will depend on how large the dilaton mass is.

Notice that in \rnexp\ there is only one free parameter, since we are
looking for an asymptotically flat solution with a particular charge
and mass. As we approach the singularity with $Q$ fixed, there are
however four residual free parameters. Thus, since our solution space
is five-dimensional, we do in general expect a solution to exist,
however without for example a numerical integration, this is not a
certainty\footnote{$^\dagger$}{This issue is currently being
addressed by Horne and Horowitz, and we thank them for discussions on
this point.}. However, in certain cases, we can eliminate
possibilities for the causal structure of the solutions, and we will
therefore concentrate on what we can say analytically about the
general properties a potential must have to admit one, two or three
horizons.

We start by proving that the potential $V_2(\phi)$ can have at most
two horizons. To do this we integrate the
middle eqn of \massv\ between the first and second horizons.
This gives
\eqn\vbound{\int_{r_1}^{r_2} 2C^2 (\half \dvdp -V) <0.}
At first sight this may not seem at all restrictive, but for
the potential $V_2(\phi)$ we have
\eqn\restrict{\half {\partial V_2 \over \partial \phi} -V_2 =
{m^2} (1-e^{-2 (\phi-\phi_0)}).}
Since we have already shown that $\phi$ is decreasing on the
interval $[r_1,r_2]$, \vbound\
would require $\phi(r_2)<\phi_0$. But a rearrangement of \massv,
\eqn\remassv{ [A^2 C^2 \phi']' = C^2 (V + \half \dvdp) + (A^2 C C')'
-1}
integrated in a neighborhood of $r_2$ gives
\eqn\intmass{A^2 C^2 \phi' |_{r_2+\delta} = \int_{r_2}^{r_2+\delta}
{m^2 C^2} (e^{2 (\phi-\phi_0)} -1) - \delta + A^2 C C'
>0}
which implies
\eqn\intimp{
\int_{r_2}^{r_2+\delta} {m^2 C^2} (e^{2 (\phi-\phi_0)}-1) >0}
so that $\phi(r_2)>\phi_0$. So, $V_2(\phi)$ does not admit a black hole
with three horizons. $V_1$ on the other hand has
\eqn\voneb{ V_1 - \half {\partial V_1 \over \partial \phi} =
               2m^2 (\phi-\phi_0) (\phi- \phi_0-1)}
thus as $\phi-\phi_0$ becomes greater than $1$, \vbound\ can be satisfied.
This can be seen to fit in with some of our earlier intuitive
arguments. Since $V_2$ becomes very important for large $\phi$
compared to $V_1$, we might expect greater resistance to approaching a
GHS massless solution. Therefore for large dilaton mass we might
expect the solution to remain much like Reissner-Nordstr{\o}m except
very close to the singularity, by which stage there is no possibility
of a third horizon forming.

It is obviously now of interest to determine whether the solution for
$V_1$ can have three horizons, not least because of the bizarre
causal structure associated with three horizons! It would also be
useful to know whether, and if so when, even two horizons are
possible. One of the nice
features of the massless dilaton black hole was that it had a
(presumably stable) Schwarzschild like causal structure, with no
Cauchy horizons. It would therefore be useful to know if this
single horizon structure persists, and if so, for approximately
what range of parameters.

In order to get a clearer picture of what is happening, and to
simplify
some of the arguments, let us transform to
the string metric \stringg. The action in the new metric is
\eqn\confact{
S = \int d^4x \sqrt{-{\hat g}} e^{-2\phi} \left \{ -{\hat R}
- 4{\hat g}^{ab} \partial_a\phi \partial_b\phi - {\hat g}^{ac}
{\hat g}^{bd} F_{ab} F_{cd} - e^{-2\phi} V(\phi) \right \}
}
The equations of motion which follow from this action are
\eqn\pre{
8{\hat \nabla}_a {\hat \nabla}^a \phi - 8 ({\hat \nabla} \phi)^2
+ 4 e^{-2\phi} V(\phi) = e^{-2\phi} {\partial V \over \partial\phi}
- 2 {\hat R} - 2F_{ab}^2
}
\eqn\prer{
{\hat R}_{ab} + 2 F_{ac}F_b^{~c} + 2 {\hat \nabla}_a
{\hat\nabla}_b\phi
= \half {\hat g}_{ab} \left \{ {\hat R} + F_{ab}^2 +
e^{-2\phi}V(\phi)
+ 4{\hat\nabla}_a {\hat\nabla}^a \phi - 4 ({\hat\nabla}\phi)^2 \right
\}
}
where all contractions are taken with the new string metric, ${\hat
g}_{ab}$.
Taking the trace of \prer\ simplifies \pre:
\eqn\fieq{
-4{\hat\nabla}_a {\hat\nabla}^a \phi + 8 ({\hat\nabla}\phi)^2 =
e^{-2\phi} {\partial V \over \partial\phi} - 2 F_{ab}^2
}
and \prer\ then can be written as
\eqn\stmetric{
{\hat G}_{ab} = -2 {\hat\nabla}_a {\hat\nabla}_b \phi - 2 F_{ac}
F_b^{~c}
+ \half {\hat g}_{ab} [ 3 F_{cd}^2 + e^{-2\phi} V(\phi)
- e^{-2\phi} {\partial V \over \partial\phi}
+ 4 ({\hat\nabla}\phi)^2 ]
}
The Maxwell field equation is unchanged, hence $F = Q\sin\theta
d\theta \wedge d\phi$ is also a solution in this theory.

Looking for a static spherically symmetric solution as before
with metric
\eqn\hatmet{
d{\hat s}^2 = {\hat A}^2 d\tau^2 - {\hat A}^{-2} d\rho^2 - {\hat C}^2
\{ d\theta^2 + \sin^2 \theta d\phi^2 \}
}
gives for the dilaton
\eqn\newdil{
[{\hat A}^2 {\hat C}^2 (e^{-2\phi})']' = {2Q^2 e^{-2\phi} \over {\hat
C}^2} -
\half {\hat C}^2 e^{-4\phi} {\partial V\over \partial\phi}
}
and the equations of motion for the metric variables can be boiled
down to
\eqn\hateq{
\eqalign{
{\hat C}'' &= {\hat C} \phi'' \cr
[ ({\hat A}^2)' {\hat C}^2 e^{-2\phi} ]' &= - {\hat C}^2 e^{-4\phi}
(V- \half {\partial V \over \partial\phi}) \cr
[ {\hat A}^2 ({\hat C}^2)' e^{-2\phi} ]' &= e^{-2\phi} \left [
2 - {4Q^2 \over {\hat C}^2} - {\hat C}^2 e^{-2\phi}
(V-\half {\partial V \over \partial\phi}) \right ] . \cr
}}

Note immediately that the last two `gravity' equations imply
\eqn\censor{
[({\hat A}^2 ({\hat C}^2)' - ({\hat A}^2)' {\hat C}^2 ) e^{-2\phi}]'
= \left ( 2 - {4Q^2 \over {\hat C}^2} \right ) e^{-2\phi}
}
Thus if ${\hat C}^2 > 2Q^2$ for all $\rho$, then $({\hat A}^2 ({\hat
C}^2)'
- ({\hat A}^2)' {\hat C}^2 ) e^{-2\phi}$ must be monotonically
increasing;
this is not compatible with a third inner horizon.
In fact, \censor\ shows that if there are to be three horizons
not only is $\hat C < 2Q$ required, but also a turning point in $\hat
C$
before the final inner horizon is reached.
Note that although $C(r)$ in the original metric
is strictly increasing, this does not imply a similar result for
${\hat C}
(\rho)$, since
\eqn\area{
\eqalign{
{d\over d{\hat s}} {\hat{\cal A}}(\hat s) &= 8\pi{\hat C}
|{\hat A}^2|^{1/2} {\hat C}'(\rho) \cr
&= e^{-\phi} {d\over ds} (e^{2\phi} {\cal A}) =
8\pi C |A^2|^{1/2} e^\phi ( C'(r) + C \phi'(r) ) \cr
}}
hence if $\phi'(r) < -C'/C$, ${\hat C}'(\rho)$ can be negative.
Thus in the string metric, the event horizon actually masks
undulations in the $t=$constant surfaces, areas of the
two-spheres actually increase before the innermost horizon.

Now let us examine the equations of motion for a weak potential (by
which we mean $\displaystyle{\sup_{\phi>0} 2Q^2 e^{-2\phi}V<1}$)
at a putative inner
horizon. We first note that
since $C(r)$ is an increasing  function of
$r$, $e^{-2\phi}{\hat C}^2$ is an increasing function of $\rho$. Thus
\eqn\areap{
[{\hat A}^2 ( e^{-2\phi} {\hat C}^2)']' =
\left [ 2 - {2Q^2\over {\hat C}^2}  - {\hat C}^2 e^{-2\phi} V(\phi)
\right ] e^{-2\phi}
}
implies that the term in square brackets on the r.h.s.\ of \areap\ is
negative at the inner horizon. We would like to use this to establish
an upper bound on ${\hat C}^2$. First notice that \censor\ implies
that ${\hat C}^2 < 2 Q^2$  for some range between the outer horizon
and the putative inner horizon. If we assume that ${\hat C}^2$ is
strictly increasing between the inner and outer horizon then we
also know that the previous inequality is satisfied at the inner horizon.
Thus solving \areap\ as a quadratic for $2Q^2/{\hat C}^2$ gives
\eqn\wf{
{\hat C}^2 < { 2Q^2 \over 1 + \sqrt{ 1 - 2Q^2e^{-2\phi}V}} .
}
The equation of motion for the dilaton \newdil\ implies
\eqn\waf{
{2Q^2 \over {\hat C}^2} < \half {\hat  C}^2
e^{-2\phi} {\partial V \over \partial \phi}.
}
Equations \wf\ and \waf\ thus give us
\eqn\ceein{
{2Q^2 \over \sqrt{Q^2 e^{-2\phi} \partial V /\partial \phi} }
< {\hat C}^2 < { 2Q^2 \over 1 + \sqrt{ 1 - 2Q^2e^{-2\phi}V}  }
}
which in turn requires
\eqn\ineq{
1 < {Q^2 e^{-2\phi} (V+ {1\over 2}{\partial V\over\partial\phi})^2 \over
\partial V/\partial\phi} .
}
If we are looking for a lower bound on $Qm$, then we maximize the r.h.s.\
of \ineq\ with respect to $\phi$. The maximum occurs when
\eqn\maxxi{
\left ( 2 {\partial V \over \partial \phi} + {\partial^2 V \over
\partial \phi^2} \right ) \left ( {1\over 4} \left ( {\partial V\over
\partial \phi} \right ) ^2 - V^2 \right ) = 0 .
}
If the potential is convex, ${\partial V\over \partial\phi}$ will be
positive for positive $\phi$, hence the maximum will occur when
\eqn\limiter{
V(\phi_{\rm max}) = \half {\partial V \over \partial \phi} \Biggl |
_{\phi_{\rm max}} \;\;\;\;\;\; \Rightarrow \;\;\;\;\;\;\; 1 <
2V(\phi_{\rm max}) Q^2 e^{-2\phi_{\rm max}} .
}
Obviously this is in contradiction with our initial supposition
that the potential was weak. So, since
generically this r.h.s.\ will have order of magnitude $(Q^2m^2)$, hence
for $Qm < $O(1), there can be only one horizon. For our test
potentials, \limiter\ gives for $V_1$, $\phi_{\rm max}- \phi_0=1$,
$Qm>e^{1+\phi_0}/2$ for an inner horizon, and for $V_2$,
$\phi_{\rm max}-\phi_0$`='$\infty$ giving $Qm>e^{\phi_0}$.
Although we have found several arguments that suggest  the validity of
this reasoning, which is further supported by
the analysis of extremal solutions in the following section,
we do not have a watertight proof  that ${\hat C}'>0$.

We end this section by mentioning that the transformation
\dual\ can be used to trivially construct electrically charged
solutions from the magnetic solutions discussed above as long
as the potential $V(\phi)$ is an even function of $\phi$.

\newsec{Extremal solutions}

In thinking about the extremal limit(s) of a black hole with a
massive dilaton, the situation is more diverse than either
Reissner-Nordstr{\o}m
or massless dilaton black holes.
In these  cases, there is a unique extremal limit:
For Reissner-Nordstr{\o}m, $Q=M$ and the inner and outer horizons merge, on the
verge of disappearing and leaving a naked singularity. The
resulting Penrose-Carter diagram is shown in fig. 4.
\ifig\ffour{Penrose-Carter diagram  for a  extremal Reissner-Nordstr{\o}m black
hole.}{\epsfysize=5.4in \epsfbox{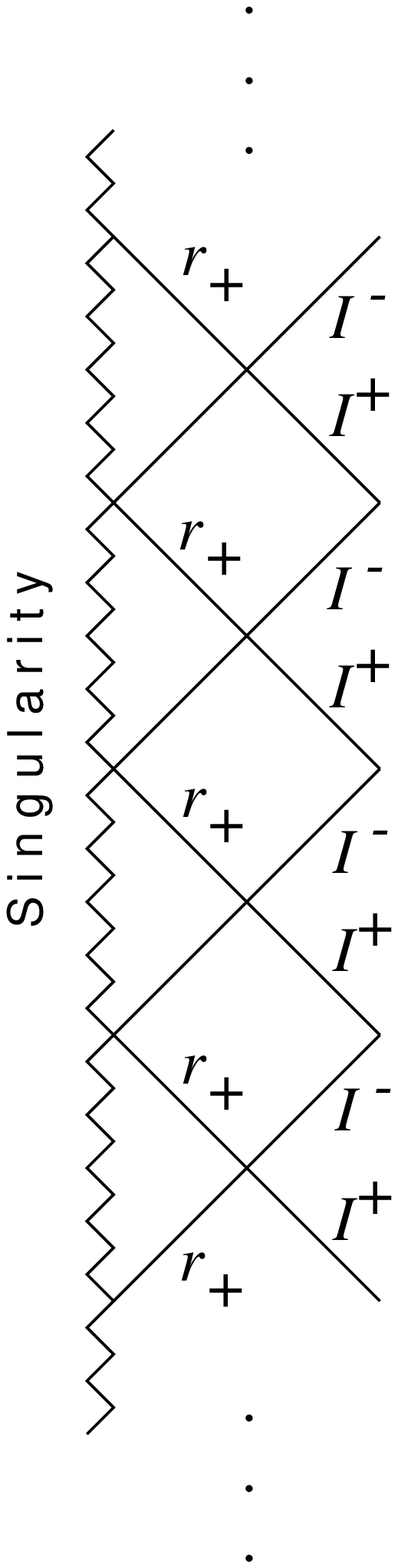}}
For massless dilaton black holes
the singularity and horizon merge, the singularity again on the
verge of becoming naked, although this time by moving `outside'
the event horizon with a Penrose-Carter diagram as
shown in fig. 2. For the case of massive dilatons however, we
have several options, depending on the number of horizons. For
example, if we have only one horizon, we might expect an extremal
limit similar to the massless case, but for $V_1=2m^2 (\phi -\phi_0)^2$
there is also
another
possibility, namely that $A^2$ develops first a stationary, then two
turning points, and finally an additional double horizon. In other
words, black holes with massive dilatons can exhibit both kinds
of extremality.
We will first look at the conditions for Reissner-Nordstr{\o}m extremality,
before
analysing the massless dilaton type of extremality.

In looking for a Reissner-Nordstr{\o}m type extremal solution, the defining
feature
is the repeated horizon. This corresponds to a repeated zero of $A^2$,
or indeed ${\hat A}^2$. In other words $A^2 = (A^2)' = 0$ at such a
point.
Examining the equations of motion in the string metric yields the
following equalities that must then be satisfied:
\eqn\exthri{
\eqalign{
2 - {2Q^2 \over {\hat C}^2 } - {\hat C}^2 e^{-2\phi} V &=0  ,\cr
{2Q^2 \over {\hat C}^2} - \half {\hat C}^2 e^{-2\phi} {\partial
V\over \partial \phi} &= 0. \cr
}}
These equalities can be solved for ${\hat C}^2$ and give
\eqn\trio{
{2Q^2 \over 1 \pm \sqrt{ 1 - 2Q^2 e^{-2\phi} V} } =
{2Q^2 \over \sqrt{ Q^2 e^{-2\phi} {\partial V \over \partial \phi}}}
}
which can then be solved for $\phi$, giving
\eqn\ineqtwo{
\partial V/\partial\phi = Q^2 e^{-2\phi} (V+ {1\over 2}{\partial
V\over\partial\phi})^2 .
}
We will examine each test
potential in turn.

For $V_1$, these boil down to a modified cubic for $\phi$
\eqn\veel{
(\phi -\phi_0)(\phi -\phi_0+1)^2 e^{-2\phi} = 1/m^2 Q^2
}
and for $V_2$
\eqn\veet{
e^{-2 \phi_0}\tanh (\phi - \phi_0) = 1/m^2Q^2 .
}
The former relation generically has two solutions for $\phi$, whereas
the latter relation has only one positive-$\phi$ solution. This is
of course because there can only be one possible extremal type for
the $\sinh^2\phi$ potential, namely a positive-$A^2$
Reissner-Nordstr{\o}m, since it does
not admit three horizons.
It is therefore appropriate that in the limiting value for $Qm=e^{\phi_0}$,
it also corresponds to $\phi - \phi_0=\infty$, {\it i.e.}~the singularity.
For $V_1$, $(\phi -\phi_0)(\phi - \phi_0+1)^2e^{-2\phi}$ has a maximum of
$4 e^{-2 \phi_0}/e^2$ at
$\phi -\phi_0=1$, therefore at the limiting value $Qm = e^{\phi_0}e/2$, the two
roots
of \veel\ coincide at $\phi -\phi_0 =1$. At such a point $(A^2)''$ also
vanishes, and there is a triple horizon.

In Einstein gravity the throat region of the extreme
Reissner-Nordstr{\o}m metric is described by an exact solution of
the Einstein equations with constant radius two-spheres sometimes
known as the Bertotti-Robinson electromagnetic universe \BR.
A similar situation prevails here. Looking for a solution of
\newdil\ -- \hateq\ with $\hat C$ and $\phi$ constant we find
a solution with $\phi$ given by \veel\ or \veet\ and with
\eqn\ghbr{\eqalign{ {\hat C}^2 &= {2 Q e^{\phi} \over \sqrt{ \partial V /
                                     \partial \phi}} \cr
                     ({\hat A}^2 )'' &= - e^{-2 \phi} (V - \half
		      {\partial V \over \partial \phi}) \cr }}
where the r.h.s. of  \ghbr\ is to be evaluated
at the solution of \veel\ or \veet\ depending on the choice of potential.

For the other type of extremal solution, the defining feature is that the
singularity and horizon coincide, in other words that $A^2=0$ at
the singularity. Searching for an expansion of the solution near the
singularity reveals an interesting difference between the massless and
massive cases, which is reflected in the two-dimensional theory as we
will see in the next section. In the massless case the solution in the
neighborhood of the singularity was a linear dilaton vacuum (LDV),
{\it i.e.}~$\phi = -\alpha\rho$, with (vanishing) corrections of the form
$e^{\alpha\rho}$. As we will see, the form of these corrections may
alter, although the LDV will still persist.
Since this `throat' structure of the
original massless dilaton black holes was so attractive for hiding
information, it
is important to demonstrate that this structure remains.

We are looking for a solution of the form
\eqn\zeroa{
\eqalign{
{\hat A}^2 &= a_1 + a_2(\rho) e^{\alpha \rho}  + \cdots\cr
{\hat C} &= c_0 + c_1(\rho) e^{\alpha \rho} + \cdots \cr
e^{-2\phi} &= f_0 e^{\alpha \rho} + f_1(\rho) e^{2\alpha\rho} + \cdots \cr
}
}
as $\rho \rightarrow \infty$.
Using ${\hat C}'' = {\hat C}\phi''$ readily shows that $f_1 = -2 f_0
c_1(\rho) /c_0$, independent of whether there is a potential or not,
however the other two equations in \hateq\ rapidly show that while
$c_0^2 = 2Q^2$ as with the zero mass case, the situation for the
other variables is quite different. The corrections, $c_1$ and $a_1$
must now satisfy non-trivial differential equations
\eqn\correct{
\eqalign{
[(a_2 e^{\alpha\rho})'e^{\alpha\rho}]' &= - f_0 e^{2\alpha\rho}
(V - \half {\partial V \over \partial \phi}) \cr
[(c_1 e^{\alpha\rho})' e^{\alpha\rho}]' &= {1\over 2Q^2 a_1}
e^{2\alpha\rho} [ 4c_1 - 2\sqrt{2} Q^3 f_0 (V-\half {\partial V\over
\partial \phi}) ] \cr
}}
and
\eqn\aone{
2Q^2 \alpha^2 a_1 = 1 - Q^2 \left [ e^{-2\phi} {\partial V\over
\partial \phi} \right ]_{\phi = -\half (\alpha\rho + \ln f_0)}
}
These equations can be solved for one's chosen potential. For example,
for $V_1 = 2m^2(\phi-\phi_0)^2$, $a_1 = e^{\phi_0}/2Q^2\alpha^2$,
as in the zero mass case, but the corrections take the form
\eqn\vlcorr{\eqalign{ a_2 &= - {f_0 m^2\over 16\alpha^2} \left[ 2\rho^2
\alpha^2- 2\rho\alpha
(1-2\ln f_0) + 1 - 2\ln f_0 + 2\ln^2 f_0 \right] \cr
c_1 &= - {m^2 Q^3 e^{-3\phi_0} f_0 \over 162\sqrt{2}}
\Biggl[ 27 \alpha^3 \rho^3 +
27 \alpha^2\rho^2 ( 2 + \ln f_0) \cr & \qquad - (3 \alpha\rho - 1)
(4 - 16\ln f_0 - 9\ln^2 f_0) \Biggr]  + K/3\alpha \cr
}}
where $K$ is an integration constant.

For $V_2$ the result is much simpler: $a_1 = (e^{2\phi_0} -
Q^2m^2)/2Q^2\alpha^2$, $a_2 = m^2 f_0 /2\alpha^2$, and
\eqn\ceel{
c_1 = -f_0 Qe^{-\phi_0}/\sqrt{2} + K e^{ \kappa\rho}
}
where $\kappa$ is a root of $k^2 + 3\alpha k -
2\alpha^2m^2Q^2/(e^{2\phi_0}-m^2Q^2) = 0$.
The main thing to note about this solution is that if
$Qm>e^{\phi_0}$, then $a_1 <0$. In other words, down the throat of the
black hole, space and time would actually reverse r\^oles. Clearly
by continuity, this can only happen if there is an event horizon
(${\hat A}^2 <0$ near the singularity implies ${\hat A}^2$ has a
zero {\it before} the singularity), which clearly means that this is
not a GHS extremal solution, where the singularity is on the verge of
becoming naked. However, it can be an  extremal solution in the sense
of a transition between two and one horizons, where the singularity
and the inner horizon merge in the interior of the black hole. In this
case, therefore, the ``throat'' of this inner extremal solution is an
inverted LDV - space and time have swapped r\^oles. A possible Penrose-Carter
diagram for such a solution is shown in fig. 5.
\ifig\ffive{Penrose-Carter diagram  for an extremal massive dilaton
black hole where one starts from an asymptotically flat region (AF)
and approaches the linear dilaton vacuum (LDV) {\it after} passing
through an event horizon.}{\epsfysize=4.8in \epsfbox{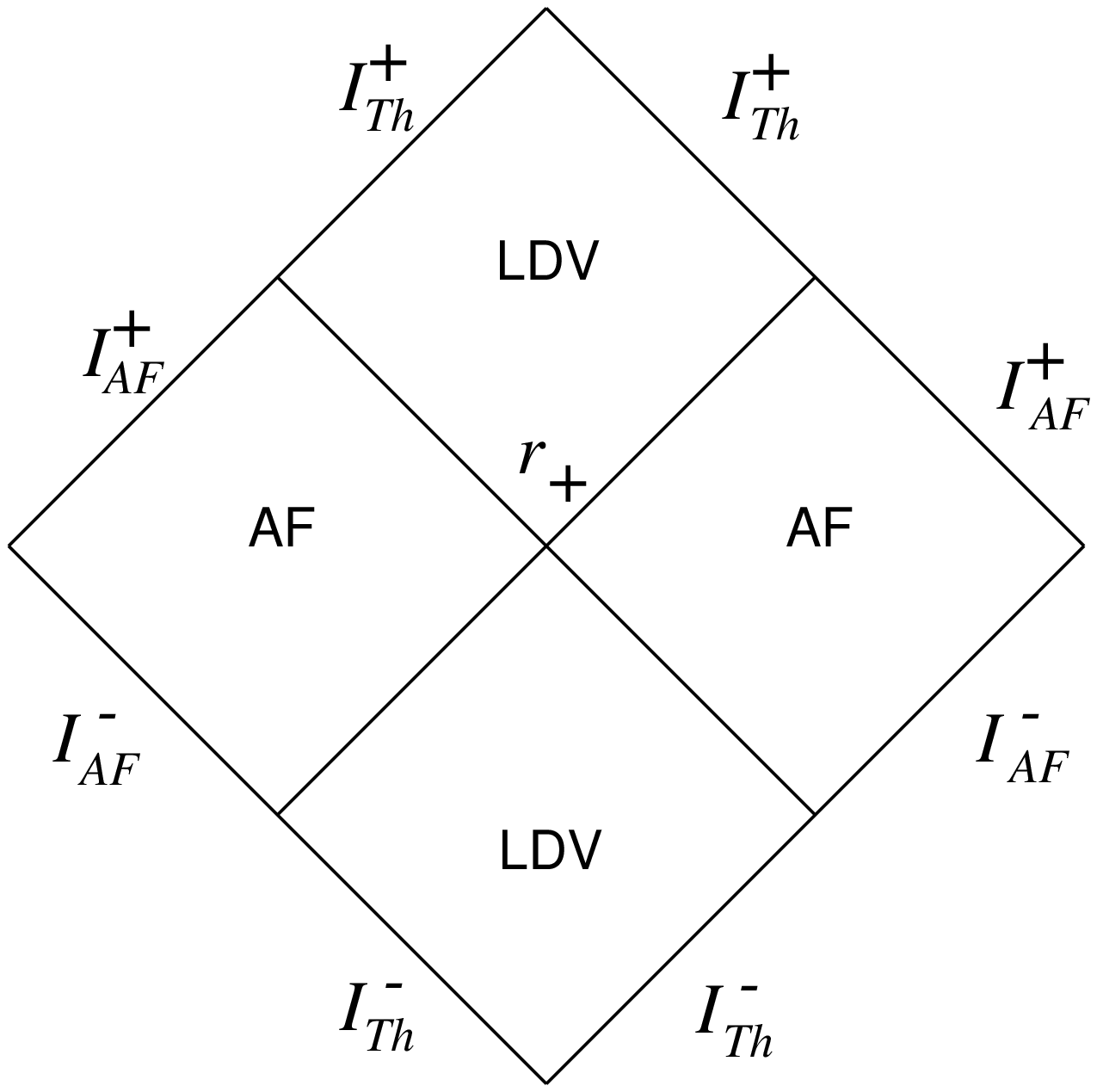}}
It is interesting
that these more massive dilatons cannot exhibit a GHS
extremal solution, which can only occur for $Qm\leq e^{\phi_0}$, the case of
equality being the combined Reissner-Nordstr{\o}m -- massless dilaton
extremal type already discussed
above. Thus the type of extremal solutions for this potential can be
quite neatly catalogued: For $Qm<e^{\phi_0}$ there can only be a massless
dilaton type extremal solution, the singularity and horizon merge.
For $Qm>e^{\phi_0}$ there can only be a Reissner-Nordstr{\o}m type
extremal solution where the singularity is about to become naked, with
also the possibility of an inverted GHS extremal solution in the
interior.  For $Qm=e^{\phi_0}$ there is a situation analogous to the
triple horizon of $V_1$, namely that the singularity merges with a
repeated horizon.


Of course we also expect some relation between $Q$, $M$, and $m$ to be
satisfied in order to obtain these various types of extremal solution,
however, unlike the massless case, we do not have an analytic
closed-form solution from which we can extract these formulae.
However, it is possible to obtain approximate information as to the
extremal mass-charge relationships.

For $mQ \gg e^{\phi_0}$, searching for a Reissner-Nordstr{\o}m
repeated horizon ($(A^2)''>0$), both \veel\ and \veet\ give the same
result, namely
\eqn\setphi{
(\phi - \phi_0) = e^{2\phi_0} /m^2Q^2.
}
Using \massv\ and \mdilly, this implies
\eqn\ceee{
C^2 = Q^2 e^{-2\phi_0} ( 1 - {1\over m^2Q^2} )
}
therefore
\eqn\asymass{
\left [ (A^2)' C^2 \right ] ^\infty _{r_H} = 2M =
\int_{r_H}^\infty \left ( {2Q^2 e^{-2\phi} \over C^2} - C^2 V \right )
< {2Q^2 e^{-2\phi_0} \over C_H}
}
Knowing that $mM \gg e^{\phi_0}$, we can use \rnexp\ to
find the exact form of the corrections which turns out to be
\eqn\realm{
M = Qe^{-\phi_0} \left ( 1 - {1\over 10m^2 Q^2
e^{-2\phi_0}} \right )
}
Thus, perhaps not surprisingly, the effect of the dilaton is to
increase the charge carried by an extremal solution. As $mQ$
decreases, this discrepancy obviously increases, so we will estimate
it at presumably what is its maximum - the triple extremal solution.

For $V_1$, this occurs at $( \phi - \phi_0) = 1$, $C^2 = Q^2
e^{-\phi_0}/ em = 1 /2m^2$, $Qm=e^{1+\phi_0}/2$.
We then use
\eqn\useful{
((A^2)'C^2)' < {2Q^2 e^{-2\phi} \over C^2} < {2Q^2 C' e^{-2\phi_0}
\over C^2}
}
integrated between the horizon and infinity to conclude
\eqn\smallm{
M  < {e\over\sqrt{2}}Qe^{-\phi_0}  = {e^2\over 2 \sqrt{2}m}
}
Obviously, this is not a concrete result, nonetheless, it will give
the correct order of magnitude for $M$. It is interesting to note
that while this looks similar to the massless black hole charge-mass
ratio when expressed in terms of $Q$, when expressed in terms of the
dilaton mass the inequality is more eloquent. It shows that unless the
dilaton mass is very small, only mini-black holes are capable of attaining
this special extremal limit. For example, if the dilaton acquires a
mass of around $1$ TeV, then the extremal black hole would have to be
{\it no heavier} than about $10^{10}g$ while if the dilaton mass is
$10^{18}$ GeV then the black hole mass would have to be less than
the Planck mass. Since the dilaton mass could lie anywhere in this range,
such solutions would be relevant only for primordial black holes.

For GHS-type extremal solutions, by integrating the final two
equations of \massv, we find
\eqn\massghs{
M = {2\over 3} \int_{r_{\rm sg}}^\infty {Q^2 e^{-2\phi}
\over C^2}
}
estimating this integral as $\displaystyle{\lim_{r\to r_{\rm sg}}}
{Q^2 e^{-\phi_0}\over C}$ gives for $V_1$, $M\sim \sqrt{2}Q/3$. It must be
stressed that this is only an estimate, therefore we should not compare
its numerical value to that of the massless extremal limit, or indeed
Reissner-Nordstr{\o}m, however, it does show that the
extremal mass-charge relationship is in the same ball park as these
other two cases.
Without a numerical solution and integration however, we cannot
be more specific. For $V_2$ this estimate gives $M\sim
\sqrt{2}Qe^{\phi_0}/3$, again, roughly the same order of magnitude.

\newsec{Two Dimensions}

Addition of a dilaton
potential
no longer allows exact solutions of the form $S^2 \times {\cal
M}^2_{BH}$
with constant radius $S^2$ and a 2-d black hole spacetime ${\cal
M}^2_{BH}$.
Nonetheless it is of some interest to study black holes in
two-dimensional
massive dilaton gravity even if there is no longer a direct
connection with four-dimensions. Solutions to two-dimensional
dilaton gravity with a dilaton potential have also been studied
in \nappi.

Motivated by the four-dimensional Lagrangian \confact\ we take as our
starting
action
\eqn\twoact{S = {1 \over 2 \pi} \int d^2 x \sqrt{-g} e^{-2 \phi}
\left(
R + 4 (\nabla \phi)^2 + 4 \lambda^2 - e^{-2 \phi} V(\phi) \right)}
Note that in this section we change our conventions to comply with
those commonly used in two-dimensional dilaton gravity (metric
signature
$(-,+)$ etc.). The equations of motion following from \twoact\ are
\eqn\twodeq{
2 e^{-2 \phi} \left[ \nabla_\mu \nabla_\nu \phi + g_{\mu \nu}
( (\nabla \phi)^2 - \nabla^2 \phi - \lambda^2  +{1 \over 4} e^{-2
\phi} V(\phi)) \right] =0,}
\eqn\twodeqt{
e^{-2 \phi} \left[R + 4 \lambda^2 + 4 \nabla^2 \phi - 4 (\nabla
\phi)^2
+ e^{-2 \phi} (\half {\partial V \over \partial \phi} - 2 V(\phi))
\right]= 0,}
with \twodeq\ resulting from variation of the metric and \twodeqt\
from
variation of the dilaton.

Looking for static solutions in a ``Schwarzschild'' gauge with
\eqn\schg{ ds^2 = -A^2(\sigma) d \tau^2 + A^{-2}(\sigma) d \sigma^2}
the dilaton equation becomes
\eqn\twodil{ -(A^2)'' + 4 \lambda^2 + 4 (A^2 \phi')' - 4 A^2
(\phi')^2 +
e^{-2 \phi} (\half { \partial V \over \partial \phi} - 2 V) =0 }
while the metric equation and constraints may be written as
\eqn\pdoub{\eqalign{ \phi'' &=0, \cr A^2 (\phi')^2 - A A' \phi'
-\lambda^2
+ {1 \over 4} e^{-2 \phi} V(\phi) &= 0 \cr }}
Now $\phi''=0$ implies $\phi= p_0 + p_1 \sigma$ and if $p_1 \ne 0$ we
can choose
$p_0=0$ by shifting $\sigma$. There are thus two cases to consider,
$\phi=p_1 \sigma$, or $\phi=p_0$.

We first consider solutions with $\phi$ constant. As is clear from
the
second equation of \pdoub, there are no such solutions when $V=0$.
With $\phi$ constant and choosing the potential to be $V(\phi)=m^2
\phi^2$
the equations reduce to
\eqn\pval{ e^{-2 p_0} p_0^2 = 4 \lambda^2 /m^2}
and an equation that says that the curvature is constant:
\eqn\rval{ R \equiv -(A^2)'' = 4 \lambda^2 (1-1/p_0).}

The function $p_0^2 e^{-2 p_0}$ has a minimum at $p_0=0$, a maximum
at
$p_0=1$ where it equals $e^{-2}$, and approaches $+\infty$ as $p_0
\rightarrow
- \infty$ and $0$ as $p_0 \rightarrow +\infty$. Thus \pval\ has one
solution
if $4 \lambda^2 /m^2 > e^{-2}$ with $R>0$; two solutions if $4
\lambda^2 /m^2
=e^{-2}$ with $R>0$ at the solution with $p_0<0$ and $R=0$ at the
solution
with $p_0=1$; and three solutions if $4 \lambda^2/m^2 <e^{-2}$ with
$R>0$ for $p_0<0$, $R<0$ for $0<p_0<1$, and
$R>0$ for $p_0>1$. These solutions are the two-dimensional analog
of the throat solutions \ghbr\ discussed in the previous section.

We next look for ``linear dilaton'' solutions with $\phi=p_1 \sigma$.
The equations \twodil\ and \pdoub\ then reduce to
\eqn\twored{\eqalign{-(A^2)'' + 4 \lambda^2 + 4 p_1 (A^2)' - 4 A^2
p_1^2
+ e^{-2 \phi}
(\half { \partial V \over \partial \phi} - 2 V)|_{\phi= p_1 \sigma}
&=0 \cr
4 A^2 p_1^2 - 2 (A^2)' p_1 - 4 \lambda^2 +
e^{-2 \phi} V(\phi)|_{\phi= p_1 \sigma} &=0 \cr } }
When $V=0$ these equations have a two-dimensional black hole solution
\tbh\ given by
\eqn\twobh{\eqalign{ \phi &= -\lambda \sigma \cr A^2 &=
               1 -2M e^{-2 \lambda \sigma} \cr } }
with $M$ the (arbitrary) mass of the black hole.

With $V \ne 0$ adding the two equations in \twored\ gives
\eqn\igiveup{ (A^2)'' -2p_1 (A^2)' = e^{-2 \phi} (\half { \partial V
\over \partial \phi} -  V)|_{\phi= p_1 \sigma} = m^2 e^{-2 p_1
\sigma} (p_1 \sigma -
p_1^2 \sigma^2) }
for the potential $V(\phi) = m^2 \phi^2$. This is easily integrated
to give (using \twored\ as well)
\eqn\asoln{\eqalign{ \phi &= \mp \lambda \sigma \cr
A^2 &= 1 -2M e^{\mp 2 \lambda \sigma} - {m^2 \over 64 \lambda^2}
e^{\pm 2 \lambda \sigma} (8 \lambda^2 \sigma^2 \mp 4 \lambda \sigma
+1) \cr}}
with $M$ arbitrary. If we want to obtain a solution which is
asymptotically
flat at one end of our one-dimensional world we must take $M=0$.
With the usual convention that the singularity occurs at $\sigma
\rightarrow
-\infty$ we then have as a solution \asoln\ with the lower choice of
sign and $M=0$. In contrast to the usual two-dimensional black hole
of \twobh\ which has a singularity at strong string coupling
$g_s \equiv e^\phi = e^{-\lambda \sigma} \rightarrow +\infty$, this
solution
has a singularity at weak string coupling with $g_s = e^{\lambda
\sigma}
\rightarrow 0$, with the potential playing a crucial role.

The causal structure of this solution depends as before on the ratio
$\lambda^2/m^2$. In particular, the function $f(x)=e^{-2x}(8x^2
+4x+1)$
appearing in \asoln\
has a maximum at $x=(1+\sqrt{3})/4$ where it takes the value
$f_M= e^{-(1+\sqrt{3})/2}(4+2 \sqrt{3})$ and the solution has one,
two,
or three horizons depending on whether the ratio $64 \lambda^2/m^2$
is larger
than $f_M$, equal to $f_M$, or less than $f_M$ respectively.

Of course it is not clear that this choice of potential plays
any particular role in two dimensions, and one might argue that
it is not physically sensible to add a term which dominates at
weak coupling. Another choice of potential of some interest
is $V(\phi)= e^{2 \phi} m^2 \phi^2$. Repeating the previous analyses
with this potential we find two types of deSitter (constant
curvature)
solutions. If $\phi=p_0$ is constant we now find a solution with
\eqn\newdesit{\phi^2 = p_0^2 = 4 \lambda^2/m^2 , \qquad R = -(A^2)''
= -m^2 p_0 ,}
and if $\phi=p_1 \sigma$ we find a general solution
\eqn\desittwo{ A^2 = 1 - {m^2 \over 4 p_1} \sigma - {m^2 \over 4}
\sigma^2
                        - 2M e^{2 p_1 \sigma}}
with $p_1$ given by
\eqn\pone{p_1^2 = \lambda^2 ( 1 - {m^2 \over 8 \lambda^2}). }
This represents a two-dimensional black hole with a singularity
at $\sigma \rightarrow -\infty$ (for $p_1<0$) and which
asymptotically
approaches DeSitter space with constant curvature $R= m^2/2$ as
$\sigma \rightarrow + \infty$.

\newsec{Conclusions}

We have seen that an addition of a dilaton potential allows for a
richer variety of charged black hole solutions than is present either
in the case of Einstein gravity or massless dilaton gravity. Depending
on the values of the black hole mass and charge and the dilaton mass
and potential it is possible to have solutions with either one, two or
three horizons, the single horizon having a Schwarzschild structure,
the double a Reissner-Nordstr{\o}m causal structure, and the triple
horizon the causal lattice of figure 3. We were able to establish that
for our second test potential, $V_2(\phi) = 2m^2 \sinh^2
(\phi-\phi_0)^2$, only one or two horizons were possible. Our first
test potential however, $V_1 = 2m^2 (\phi-\phi_0)^2$ could possibly
have three horizons, provided $mQ$, the product of the black hole
charge and the dilaton mass, was sufficiently large.

We examined the various types of extremal solutions, which again were
more varied than either Einstein or massless dilaton gravity. We found
that there could be Reissner-Nordstr{\o}m type extrema, with two
horizons coinciding, and also GHS type extrema, with the singularity
and event horizon coinciding. In this case the causal structure in the
string metric would contain an infinite throat as in the massless
dilaton case. However,  whereas for $V_1$, the singularity always
coincided with an $(A^2)'>0$ horizon, for $V_2$, the singularity could
only be of GHS type if $Qm<e^{\phi_0}$. For $Qm>e^{\phi_0}$, the event
horizon and singularity meet in the interior of the black hole and the
throat has an inverted LDV structure leading to the Penrose-Carter diagram of
figure 5. Unfortunately, this type of extremal solution is not very
useful as a `cornucopian' since one is always doomed to travelling
down the throat, never to return to the outside world to pass on all
the information one has found; in any case, it does not even qualify
as a remnant - the Hawking temperature, $(A^2)'_{r_+} / 4\pi$, is most
definitely not zero! It might be possible for these solutions to
radiate until the outer event horizon merges with the inner
horizon/singularity, provided sufficient charge is lost, however, it
is also possible that they would turn into a Reissner-Nordstr{\o}m
extremal solution. This would depend on how preferential it was for
the black hole to discharge.

In addition to the above extremal solutions, there are also special
triple extremal solutions, where $g_{00}$ has a stationary point of
inflection. These correspond to the three horizons meeting for the
potential $V_1$, and the two horizons and singularity meeting for
$V_2$. For $V_1$ this solution has an absolute upper bound on its
mass, independent of the charge on the black hole.

Finally, we should remark that, just as in the case of massless
dilaton gravity, for every magnetic black hole solution, there is a
corresponding electric black hole solution, given by the special
duality transformation \dual.

In conclusion, while the addition of a potential destroys the
simplicity of the solution, which cannot apparently be written in
closed form, it greatly increases the wealth of the possible
spacetimes. It seems that massive dilatons allow for many more black
hole causal structures that either their massless cousins of
Einstein gravity.  It would be interesting
to further investigate these extremal solutions  since
we expect some of them to share the stability of
both extremal Reissner-Nordstr{\o}m
and massless dilaton black holes, while the presence of a dilaton
potential would seem to forbid embedding them in a theory with
unbroken supersymmetry.

\bigskip\centerline{\bf Acknowledgements}\nobreak
We acknowledge conversations with P. Bowcock, R. Geroch, J. Horne,
G. Horowitz, S. Giddings, and R. Wald. We also thank the Aspen Center for
Physics for hospitality during the course of this work.
This work was supported
in part by NSF grant PHY90-00386. R.G. was also
supported
by the McCormick Fellowship fund at the Enrico Fermi Institute; J.H.
acknowledges the support of NSF PYI Grant PHY-9157463.

\listrefs
\end